\documentclass[twocolumn]{aastex631}
\usepackage{amsmath}

\begin{document}

\title{The Photospheric Emission of a Short-Duration Gamma-Ray Burst Emerging from a Realistic Binary Neutron Star Merger}

\author[0009-0008-1678-2787]{Nathan Walker}

\author[0000-0002-9190-662X]{Davide Lazzati}
\affiliation{Oregon State University \\
Department of Physics, 301 Weniger Hall, Oregon State University\\
Corvallis, OR 97331, USA}
\author[0000-0002-4299-2517]{Tyler Parsotan}
\affiliation{Astrophysics Science Division, NASA Goddard Space Flight Center,Greenbelt, MD 20771, USA}
\author[0000-0002-3635-5677]{Rosalba Perna}
\affil{Department of Physics and Astronomy, Stony Brook University, Stony Brook, NY 11794-3800, USA}
%\affiliation{Oregon State University \\
%Department of Physics, 301 Weniger Hall, Oregon State University\\
%Corvallis, OR 97331, USA }

%% Note that the \and command from previous versions of AASTeX is now
%% depreciated in this version as it is no longer necessary. AASTeX 
%% automatically takes care of all commas and "and"s between authors names.

%% AASTeX 6.31 has the new \collaboration and \nocollaboration commands to
%% provide the collaboration status of a group of authors. These commands 
%% can be used either before or after the list of corresponding authors. The
%% argument for \collaboration is the collaboration identifier. Authors are
%% encouraged to surround collaboration identifiers with ()s. The 
%% \nocollaboration command takes no argument and exists to indicate that
%% the nearby authors are not part of surrounding collaborations.

%% Mark off the abstract in the ``abstract'' environment. 
\begin{abstract}
The almost simultaneous detection of GRB170817A and GW170817 ushered in nearly a decade of interest in binary neutron star mergers and their multi-messenger signals, resulting in a greater understanding of the processes that produce short-duration gamma-ray bursts and gravitational waves. However, open questions remain regarding the emission mechanism of these bursts.  In this work we present results from the first study of an electromagnetic signal produced from a realistic treatment of a binary neutron star merger, both for on-axis and off-axis observations. We accomplish this by using the PLUTO hydrodynamical code to inject a relativistic jet into the ejecta of a realistic binary neutron star merger, which was itself obtained from the simulation of a 3D BNS merger. Then, we model the prompt photospheric emission that would emerge from this jet using the MCRaT radiative transfer code.  We find that the resulting photon spectra can peak around $\sim 1$ MeV for on-axis emission and falls off noticeably for off-axis observations.  We also find distinctly non-thermal low and high-energy tails in multiple observations, ranging from shallow to mid-off axis observations.  Our on-axis results are consistent with the Amati Correlation for short bursts, with some strain evident at higher observing angles.  Finally, we find that the radiative efficiency is much lower than seen in previous studies of the photospheric emission of long-duration gamma-ray bursts.
\end{abstract}

%% Keywords should appear after the \end{abstract} command. 
%% The AAS Journals now uses Unified Astronomy Thesaurus concepts:
%% https://astrothesaurus.org
%% You will be asked to selected these concepts during the submission process
%% but this old "keyword" functionality is maintained in case authors want
%% to include these concepts in their preprints.

%% From the front matter, we move on to the body of the paper.
%% Sections are demarcated by \section and \subsection, respectively.
%% Observe the use of the LaTeX \label
%% command after the \subsection to give a symbolic KEY to the
%% subsection for cross-referencing in a \ref command.
%% You can use LaTeX's \ref and \label commands to keep track of
%% cross-references to sections, equations, tables, and figures.
%% That way, if you change the order of any elements, LaTeX will
%% automatically renumber them.
%%
%% We recommend that authors also use the natbib \citep
%% and \citet commands to identify citations.  The citations are
%% tied to the reference list via symbolic KEYs. The KEY corresponds
%% to the KEY in the \bibitem in the reference list below. 

\section{Introduction} 

The role of multi-messenger astrophysics in uncovering properties of black holes, neutron stars, and their associated dynamics can hardly be overstated.  Information that can be gleaned from the complimentary nature of electromagnetic radiation, gravitational waves (GWs), neutrinos, and cosmic rays from these sources can provide dramatically more insight than any one source could on its own, e.g. the time delay between GWs and electromagnetic radiation \citep{abbott_gw170817_2017,zhang_delay_2019}.  As a result, there has been considerable interest in understanding multi-messenger signals such as GWs from compact binary mergers \citep{abbott_gravitational_2017}, neutrinos from core-collapse supernovae \citep{burrows_neutrinos_1987} and active galactic nuclei \citep{the_icecube_collaboration_multimessenger_2018}, and gamma-ray bursts (GRBs) \citep{klebesadel_observations_1973,costa_discovery_1997,groot_search_1998}.

This interest has been fueled by the detection of GW170817/GRB170817A \citep{abbott_multi-messenger_2017,goldstein_ordinary_2017}, with what is so far the only observed joint GRB/GW detection.  In addition to providing smoking gun evidence of a short-duration gamma-ray burst (SGRB) associated with a binary neutron star (BNS) merger \citep{abbott_gw170817_2017,abbott_multi-messenger_2017}, UV and optical follow-up observations revealed a blue kilonova in the electromagnetic counterpart to the BNS merger \citep{nicholl_electromagnetic_2017,troja_luminous_2018,smartt_kilonova_2017,arcavi_optical_2017,chornock_electromagnetic_2017,coulter_swope_2017,drout_light_2017,kasliwal_illuminating_2017,pian_spectroscopic_2017,soares-santos_electromagnetic_2017,tanaka_kilonova_2017,valenti_discovery_2017,diaz_observations_2017,utsumi_j-gem_2017}. The confluence of gravitational waves (GWs), an SGRB, and a kilonova from this one event makes GW170817/GRB170817A an ideal test bed for multi-messenger astrophysics.

In the years since this momentous discovery, the subsequent observation runs carried out by LIGO, VIRGO, and KAGRA, O3 and O4, have failed to provide further joint GRB/GW detections. This lack of joint detections has increased the relative importance of GW170817, which has led the community to extract as much information from this event as possible .  These efforts include the identification of GRB170817A being produced from an off-axis observation, as well as inferences of the central engine properties of GRB170817A \citep{gottlieb_cocoon_2018,lazzati_intrinsic_2020,lazzati_late_2018-1,zhang_delay_2019,salafia_accretion--jet_2021,hamidani_jet_2020}. Thus, GW170817/GRB170817A remains a crucial source for testing models of the various astrophysical systems that produce GWs and SGRBs. Here we apply state-of-the-art computational methods to model, in considerable detail, the initial burst of high-energy electromagnetic radiation, the so-called prompt emission, of an SGRB. 

The GRB prompt emission exhibits a diverse temporal morphology (e.g. \citealt{fishman_gamma-ray_1995}), and in SGRBs it can last for $\sim 2$ s or less \citep{poolakkil_fermi-gbm_2021}. This emission is thought to originate in an ultra-relativistic jet launched by the GRB central engine \citep{kumar_physics_2015,piran_physics_2005}, which for GRB170817A and possibly other SGRBs, is either a magnetar or black hole remnant of a binary neutron star merger \citep{abbott_multi-messenger_2017}. Beyond this, however, the detailed radiative processes that produce and shape the prompt emission, which are critical for extracting as much information as we can from GRB170817A, have eluded astronomers in the decades since the initial discovery of GRBs in the late 1960s.   

Currently, there are two leading models that explain the prompt emission of GRBs. One of them, the synchrotron shock model (SSM), explains prompt emission by attributing observed spectra to photons emitted from relativistic electrons gyrating in a high-intensity magnetic field originating from internal shocks \citep{rees_unsteady_1994,zhang_internal-collision-induced_2010,lyutikov_gamma_2003}. Because this radiation is produced after the outflow has reached the photosphere, its spectrum would not be affected by propagation and would retain its synchrotron character. The SSM has the benefit of being relatively simple from a physical perspective and is able to fit the spectra of a number of bursts (e.g. \citealt{burgess_gamma-ray_2019}).  However, there are certain spectral features that cause strain with observations \citep{beloborodov_regulation_2013}, and, perhaps most critically, it cannot naturally reproduce various observational correlations deduced from a large set of long and short GRBs \citep{amati_intrinsic_2002,yonetoku_gamma-ray_2004,zhang_internal-collision-induced_2010,golenetskii_correlation_1983}.

On the other hand, photospheric models of GRB emission assume that radiation is produced early on in the jet history, when it is very hot and dense near the central engine \citep{peer_observable_2006,giannios_spectral_2007,lazzati_very_2009,beloborodov_collisional_2010,ryde_observational_2011}.  Thus, photoshperic models of prompt emission assume that the radiation is initially thermal.  However, processes such as sub-photospheric dissipation and multi-color blackbody emission \citep{chhotray_gamma-ray_2015,ito_monte_2018,parsotan_photospheric_2018,peer_theory_2011} can shape this thermal radiation into the characteristic non-thermal spectra of GRBs. Photospheric models are able to reproduce many features of GRB spectra, while exhibiting strain with low-energy slopes \citep{beloborodov_collisional_2010}, generally producing spectra that are more thermal at low energies than are observed in GRBs.  However, photospheric models are much better at reproducing observational correlations than synchrotron models \citep{thompson_thermalization_2007,lazzati_high-efficiency_2011,fan_photospheric_2012}.

A considerable amount of work has already been done on modeling at least some aspects of photospheric GRB prompt emission, although the vast majority \citep{lazzati_monte_2016,lazzati_photospheric_2013,parsotan_photospheric_2018,parsotan_photospheric_2020,parsotan_photospheric_2021,ito_global_2021,ito_monte_2018} of this has been focused on long-duration gamma-ray bursts (LGRBs), which can last for up to $\sim 100$ s and are produced during some core-collapse supernovae.  Much of this work has been carried out using the Monte Carlo Radiation Transfer (MCRaT) code \citep{lazzati_monte_2016,parsotan_monte_2018}, which takes results from a hydrodynamical simulation and computes radiative transfer in a post-processing step. In this paper we apply MCRaT to an SGRB for the first time. We accomplish this by injecting a relativistic jet into the ejecta of a binary neutron star (BNS) merger, obtained from a simulation carried out by \cite{ciolfi_first_2019}. The initial phase of injection was carried out by \cite{lazzati_two_2021}, and here we continue this work by simulating the jet until it reaches $\sim 10^{13}$ cm. While similar methods were used by \cite{ito_global_2021}, this work is distinguished by the use of a realistic ambient medium of a binary neutron star merger obtained from \cite{ciolfi_first_2019}, a focus on resolution over domain size, and by simulating observations further off-axis than has been done previously.  

This paper is organized as follows: In Section 2 we discuss the computational set-up for both our hydrodynamic and our radiative transfer simulations.  In Section 3 we present results from both stages of our simulations, and in Section 4 we present a discussion.

\section{Computational Methods} 

\subsection{Hydrodynamics}
We conduct a special relativistic hydrodynamical (RHD) simulation of a relativistic jet propagating through the realistic ejecta of a BNS merger, using the RHD module of the open source PLUTO code \citep{mignone_pluto_2007}, which solves the conservation equations (c.f. \citealt{marti_numerical_2003})
\begin{equation}
\begin{split}
    &(\rho u^{\nu})_{;\nu}=0\\
    &T^{\mu \nu}_{\,\,\,\,\,\, ;\nu} = 0,
\end{split}
\label{eq:cons}
\end{equation}
where  $`` ;\nu"$  denotes the covariant derivative with respect to coordinate $x^{\nu}$,
\begin{equation}
    T^{\mu \nu} = \rho hu^{\mu} u^{\nu} - p g^{\mu \nu}
    \label{eq:stress}
\end{equation}
are the components of the stress-energy tensor, $u^\mu$ are the components of the four-velocity of the fluid, $h = 1 + \dfrac{ \varepsilon}{c^2} + \dfrac{p}{\rho c^2} = 1 + \dfrac{4p}{\rho c^2}$ is the specific enthalpy, and $\varepsilon$ is the specific internal energy, which for a radiation dominated flow is  $\varepsilon = \dfrac{3p}{\rho}$. 

The simulation described here is a continuation of \cite{lazzati_two_2021} (hereafter L21). In that work, the authors carried out a 3D RHD simulation of a relativistic jet propagating through the ejecta of a BNS merger, which was itself obtained from a GRMHD simulation carried out in \cite{ciolfi_first_2019}. Here, we focus on
 the relativistic dynamics of the injected jet and on the radiative processes that produce the prompt GRB emission.  Specific details regarding the BNS merger simulation can be found in \cite{ciolfi_first_2019}. 

As initial conditions we take the jet obtained from L21, which is embedded in a  domain that extends from $\theta = 0^{\circ}$ to $\theta = 90^{\circ}$, measured from the jet axis, and from $10^8$ cm out to a radius of $r_h \sim 10^9$ cm, measured from the central engine. We project this 3D jet into 2D axisymmetric spherical coordinates by averaging the hydrodynamical quantities over the azimuthal angle $\phi$, and embed it into a density profile with an exponential cutoff, chosen due to the short timescales associated with BNS mergers in comparison to collapsars\footnote{We note that our treatment does not include the possible presence of a high velocity tail of ejecta, as postulated by \cite{kasliwal_illuminating_2017}. Such a tail might be necessary for obtaining high-luminosity shock breakout emission.}. This profile has the form
\begin{equation}
    \rho (r) = \rho_{0} e^{-r/r_{0}} + \rho_{ISM},
    \label{eq:density}
\end{equation}
for $r > r_h$, with $\rho_0 = 6.18 \times 10^{6} \text{g} \, \, \text{cm}^{-3}$ , $\rho_{ISM} = 10^{-14} \, \text{g} \, \text{cm}^{-3}$, and $r_{0} = 3 \times 10^{8} \, \text{cm}$. $\rho_{ISM}$ was chosen to prevent the density from getting too small in our computational domain, while $\rho_0$ and $r_0$ were chosen to be consistent with equation 1 in L21, which was 
in turn derived from the amount of heavy ejecta 
($M_{ejecta} \sim 0.042$ $M_{\odot}$) found in the simulation of \cite{ciolfi_first_2019}. 
We used analogous expressions for the pressures $p_{0}$ and $p_{ISM}$, where for the former the local asymptotic Lorentz factor
\begin{equation}
    \Gamma_{\infty} = \Gamma \left( 1 + \frac{4p}{\rho c^{2}} \right),
    \vspace{1mm}
    \label{eq:asym_gamma}
\end{equation}
is subject to the condition that the right hand term in the parentheses is $\sim 10^{-3}$. For the latter, $p_{ISM}/c^2 = 10^{-2}\rho_{ISM}$, which differs from the expression for $p_0$ in order to prevent code stability issues.  Finally, we set $v(r) = 0$ for the ambient medium ($r>r_h$). It is worth noting that all the preceding equations assume spherical symmetry, which is a consequence of the isotropic magnetically driven outflow evident in \cite{ciolfi_first_2019}.

The simulation described in L21 followed the first 1~s of jet injection, which will need to be accounted for when performing mock observations as described in the following sub-section. Here, we continue jet injection for another $1 \, \text{s}$ by imposing boundary conditions consistent with a constant inflow across the inner radial boundary, within the jet injection region. This inflow is characterized by a luminosity of $L_{j} = 10^{50} \, \text{erg} \,\, \text{s}^{-1}$ with a Lorentz factor of $\Gamma_{j} = 5$ and an asymptotic Lorentz factor of $\Gamma_{\infty}=300$, which were chosen to be consistent with the jet injection in L21. We obtain the injection values for $\rho$ and $p$ by imposing these conditions on the energy flux
\begin{equation}
\begin{split}
    L_j &= c\int d\sigma_i \, (T^{0i} - \Gamma \rho c^2)\\ &= c\int d \sigma_i \,  \{\beta^i \Gamma^2 (\rho c^2 + 4p) - \Gamma \rho c^2\}\\ &= \Delta \Omega \, c r^2 \beta^r \, \Gamma_{} (\Gamma_{\infty} -1) \rho c^2 
\end{split}
\label{eq:luminosity}
\end{equation}
entering the computational domain radially at the inner boundary, where $\Delta \Omega = 2\pi (1-\cos\theta_j)$ and $\theta_j = 5^{\circ}$ is the half opening angle of the jet. This differs from $\theta_j = 10^{\circ}$ used in L21. This change reflects the hydrodynamical evolution of the jet, in which the heavy ejecta from \cite{ciolfi_first_2019} acts to collimate the jet from $10^{\circ}$ to $5^{\circ}$.  Maintaining an opening angle of $10^{\circ}$ at this point, after projecting the jet into 2D and re-interpolating onto a new computational grid, would have resulted in additional, un-physical shocking as the jet attempts to push back against the cocoon structure. The jet injection parameters chosen here are broadly consistent with values inferred from observations of GRB170817 \citep{troja_year_2019,lazzati_intrinsic_2020}.

After the jet injection stops, we allow the jet to propagate out to a distance of $\sim 10^{13}$~cm, which we accomplish by breaking up the RHD simulation into three stages. Initially, the jet is superimposed on a grid with an inner radial boundary of $10^8$~cm, which is consistent with the inner edge in L21, and an outer boundary of $10^{12}$~cm.  When the jet reaches $\sim 10^{11}$~cm we cut out the inner radial boundary at $10^{10}$~cm and allow the jet to propagate out to $10^{12}$ cm, at which point we expand the outer boundary out to $10^{13}$ cm and simulate the jet out to the edge of the domain. Throughout all three stages we use an angular resolution of $\Delta \theta = 0.1^{\circ}$ from $\theta = 0^{\circ}$ to $\theta = 20^{\circ}$. From $\theta = 20^{\circ}$ to $\theta = 90^{\circ}$ the grid is covered by 100 angular cells which have a grid size close to $\Delta \theta = 0.1^{\circ}$ for $\theta \gtrsim 20^{\circ}$ and which get progressively larger as $\theta$ approaches $90^{\circ}$.  In the first two stages we use a radial grid cell of $\Delta r = 3 \times 10^6$ cm, while in the last stage we increase this to $\Delta r = 1.5 \times 10^7$ cm. Even in the last stage of the RHD simulation we resolve the jet/cocoon system into $\sim 400,000$ grid cells, spanning $\sim 1 $ light-second radially out to $20^{\circ}$ from the jet axis.

\subsection{Radiative Transfer}

After the head of the jet reaches $\sim 10^{13}$ cm we compute the radiative properties of the jet using the MCRaT radiative transfer code \citep{lazzati_monte_2016}. The original algorithm underlying MCRaT is described in detail in \cite{lazzati_monte_2016}, with improvements described in \cite{parsotan_monte_2018}.

Using MCRaT, we inject photons into the hydrodynamical outflow described in the previous sub-section.  The initial properties of the injected photons are obtained from the hydrodynamical variables of the jet, assuming they arise from a blackbody spectrum. Then MCRaT individually Compton scatters each photon under conditions obtained from the hydrodynamical pressure, density, and velocity of the jet. This process is repeated, allowing the photons to propagate through and with the jet for each of the hydrodynamical frames.  In this way, MCRaT calculates the radiative properties of the jet while accounting for the complicated dynamical processes therein.  

We inject 10,000 photons distributed across $20^{\circ}$ relative to the jet axis, which allows us to conduct mock observations at larger viewing angles than any previous work. These photons are injected into the RHD simulation with a spatial resolution of $\Delta r = 1.5 \times 10^7$ cm and a temporal resolution of $\Delta t = 0.2$ s, resulting in a light crossing ratio of $c \Delta t / \Delta r \sim 1300$. This quantity describes the ability of a given MCRaT simulation to continuously probe the RHD jet properties, and a value of 1300 converges according to \cite{arita-escalante_optimizing_2023}.

We use the software package ProcessMCRaT \citep{parsotan_parsotatprocessmcrat_2021} to conduct mock observations and produce lightcurves and spectra. ProcessMCRaT places a virtual detector at a given radius, $r_{d}$, and selects photons which have a velocity in a given direction specified by $\theta_{vp} = \theta_{obs} \pm \Delta \theta_{accept}/2$, where $\theta_{obs}$ is the viewing angle of the detector and $\Delta \theta_{accept} = 1^{\circ}$ is the acceptance range. The detection time of each photon is then calculated as 
\begin{equation}
    t_{d} = t_{p} + t_{real} - t_{j},
\end{equation}
where $t_{j} = r_{d}/c$ is the time it takes for a photon emitted at the central engine to travel unimpeded to the detector, $t_{real}$ is the simulation time of the hydro frame used to conduct an observation, and $t_{p}$ is the time it takes for an MCRaT photon to travel unimpeded from its location in the final hydro frame out to the detector.  In this work we added 1 s to $t_{real}$ to account for the initial 1 s of the jet injection carried out in L21. Detection times are therefore measured with respect to the launching of the jet.  We calculate the photon travel time as 
\begin{equation}
    t_{p} = \frac{r_{d} - r_{p} \cos(\theta_{p} - \theta_{obs})}{c},
    \label{eq:time_photon}
\end{equation}
where $r_{p}$ and $\theta_{p}$ are the radial and angular coordinates of the photon position, respectively, and $\theta_{obs}$ is the viewing angle used to place the virtual detector.

After conducting mock observations, we fit the spectra obtained from ProcessMCRaT to the Band function \citep{band_batse_1993},
\begin{align}
    \begin{split}
        &N_{E}(E) = \\   &  A \left( \frac{E}{100 \text{keV}} \right)^{\alpha} \exp(-E/E_{0}), 
                    \\ & \qquad \qquad  \qquad \qquad \qquad E \leq (\alpha-\beta)E_{0} 
                    \\ & A\left[ \frac{(\alpha-\beta)E_{0}}{100 \text{keV}} \right]^{\alpha-\beta} \exp(\beta - \alpha) \left( \frac{E}{100 \text{keV}} \right)^{\beta}
                    \\ & \qquad \qquad  \qquad \qquad \qquad E \geq (\alpha-\beta)E_{0},
                    \label{eq:band}
    \end{split}
\end{align}
using the Markov Chain Monte Carlo implementation in the software package emcee \citep{foreman-mackey_emcee_2013}. We fit  the low and high energy slopes, $\alpha$ and $\beta$, the break energy $E_{0}$, and the normalization $A$ by sampling this 4 parameter space and minimizing $\chi^2$. The best fit parameters of each spectra are obtained after the Markov Chain has ran for at least $50 \tau_{ac}$, where $\tau_{ac}$ is the auto-correlation time of the chain.

\section{Results} 
\subsection{Hydrodynamics}
Figure \ref{fig:hydro} shows the the density, asymptotic Lorentz factor (equation \ref{eq:asym_gamma}), and pressure maps of our RHD simulation at different times. From top to bottom, each row shows the jet at the times $t = 0, 1, 12,$ and $326$ s, respectively. At $t=0$~s, obtained from the fiducial simulation as described in section 2, there is a low density, fast moving core surrounded by a cocoon. Note that the density at t=0 in this figure is considerably less complicated than what is displayed in  Figure~2 of L21.  This results from the azimuthal averaging we used to obtain the initial conditions in this work, in contrast to the x-y plane projection shown in L21. As the jet injection continues, the low density core bores its way through much of this material, and emerges as a highly structured jet surrounded by a cocoon, as seen at $t=1$~s when we stop the jet injection. This structure is still present at $t = 12$~s, as well as at the end of our PLUTO simulation at $t = 326$~s, where we see the relativistic jet surrounded by the escaped cocoon, similar to what was shown in \cite{hamidani_jet_2020} .

Figure~\ref{fig:hydro} also illustrates the angular structure of the jet. The density profile at $t=0$ s  shows that the low density core of the jet is contained within a full opening angle of 10$^{\circ}$, while the asymptotic Lorentz factor, $\Gamma_{\infty}$, shows that the collimation is exaggerated even further, with the fastest parts of the jet being contained within a full opening angle of 5$^{\circ}$. {While it is often the case in simulations of LGRBs that the jet opens up after break-out, here we find that the jet retains its narrow structure throughout the hydrodynamical simulation.  This is likely due to the the decaying exponential profile we adopt, which is less steep than power laws typically used in studies of LGRBs.

\begin{figure*}
    \begin{centering}
    \includegraphics[width=0.92\textwidth]{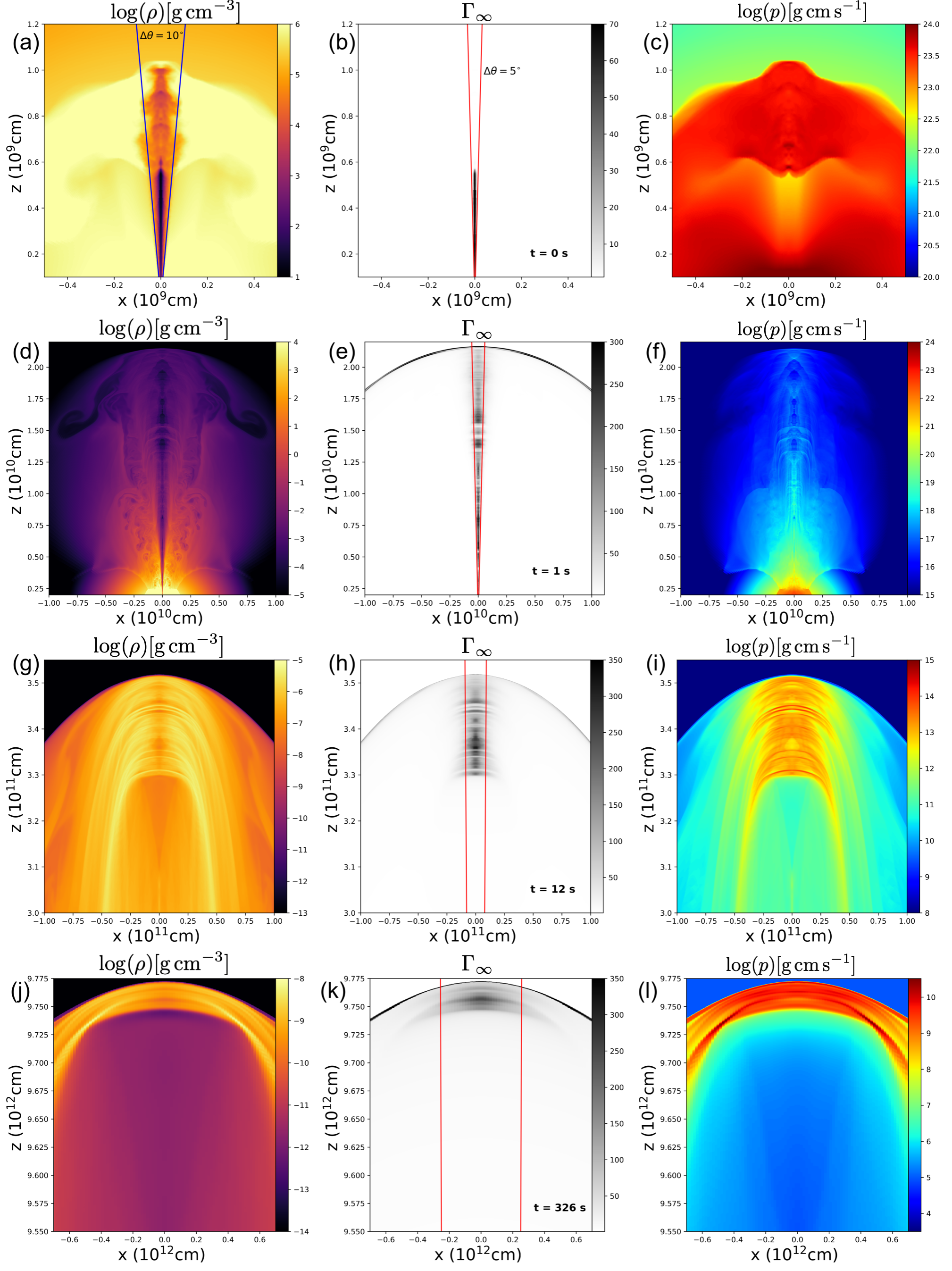}    
    \caption{Density, asymptotic Lorentz factor, and pressure plots at various times throughout our RHD simulation. The right and left columns, showing density and pressure respectively, illustrate the structure of both the relativistic jet and the mildly relativistic cocoon. In contrast, the asymptotic Lorentz factor in the middle column specifically highlights the relativistic jet. As the jet is injected and breaks out of the surrounding material, it develops and maintains a structured outflow, which is clearly visible in every panel.  In panel \textbf{(a)} the solid blue lines delineate a half opening angle of $\theta_{j} = 5^{\circ}$.  The solid red lines in each $\Gamma_{\infty}$ profile mark off a half opening angle of $\theta =2.5^{\circ}$. In panel \textbf{(k)} we see a prominent shock has developed at the front of the jet. This shock is a numerical artifact as we show in Figures \ref{fig:hydro_shock} and \ref{fig:shock_energy}.}
    \label{fig:hydro}
\end{centering}
\end{figure*}
\begin{figure*}
    
    \includegraphics[width=\textwidth]{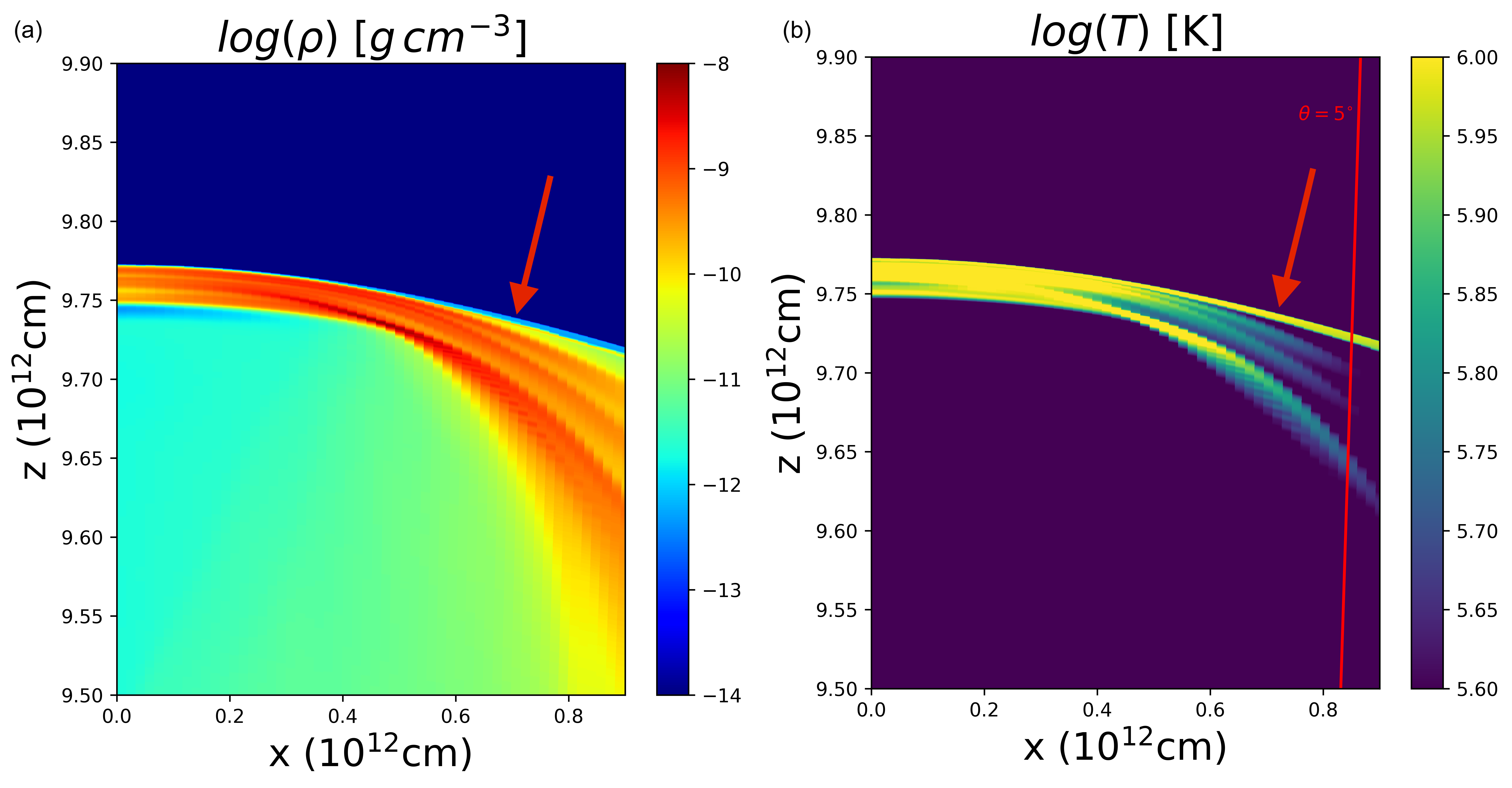}
    \caption{Pseudocolor plots of \textbf{(a)} density and \textbf{(b)} temperature at the same time as the bottom row of Figure \ref{fig:hydro}, scaled spatially to highlight the radial structure of the jet. In panel \textbf{(a)} we see the jet propagating through the interstellar medium, which is characterized with a density of  $\mathbf{\rho_{ISM} = 10^{-14}}$\textbf{ g cm}$\mathbf{^{-3}}$. Far away from the jet axis, there is a thin region just ahead of the jet that is much more dense than the surrounding medium, indicated by a red arrow in both panels. This corresponds to a region in panel \textbf{(b)} that has a much higher temperature than anything outside of the jet.}
    \label{fig:hydro_shock}
\end{figure*}

\begin{figure}
    \includegraphics[width=\columnwidth]{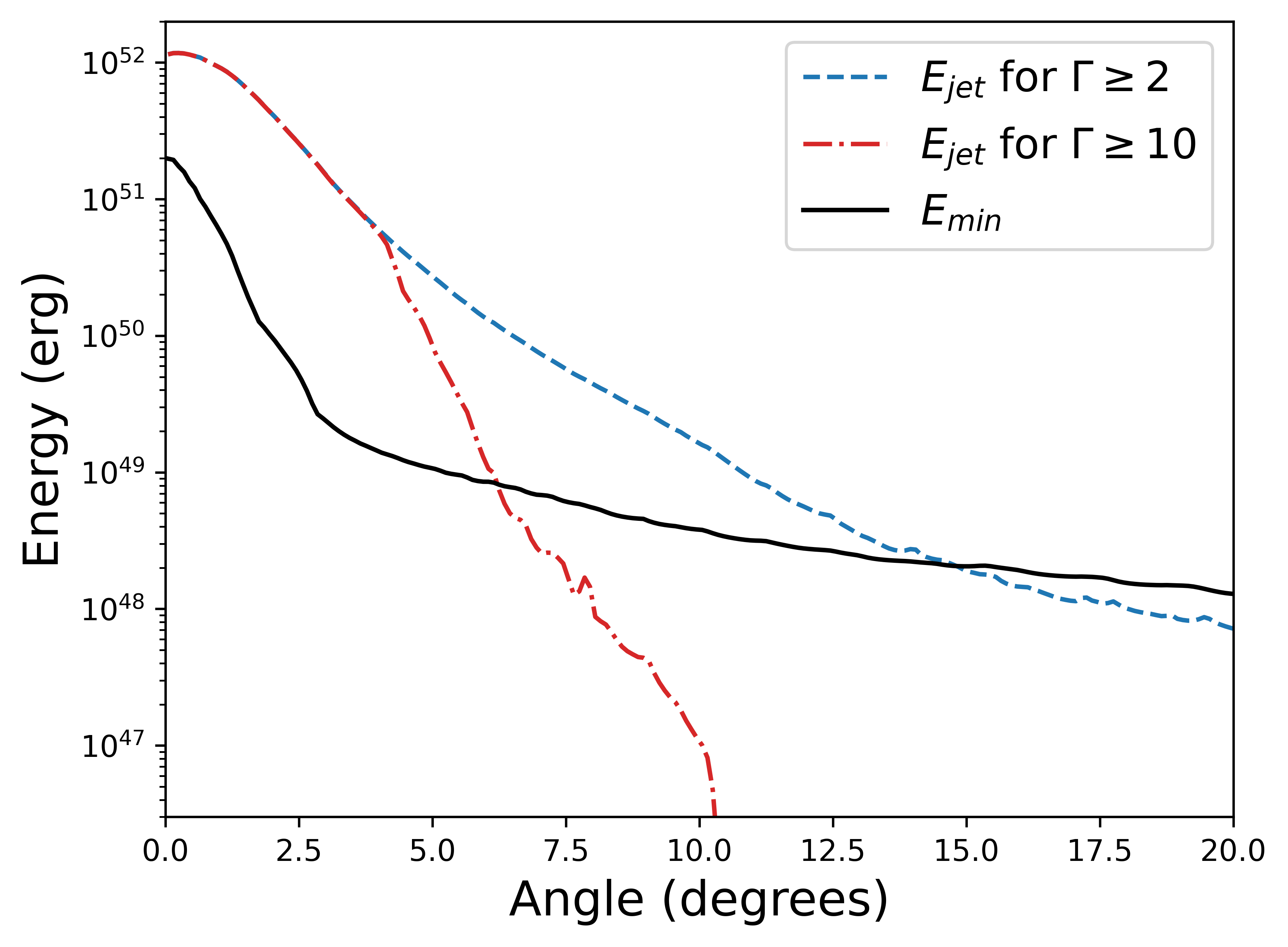}
    \caption{The energy required to form an external shock at radius $R =9.78\times10^{12}$ cm (black line), corresponding to $t=326$ s, is equal to the energy in the jet for mildly relativistic (blue dashed line) and highly relativistic (red dash-dotted line) material, as a function of angle from the jet axis.  When the energy in the jet falls below the shock energy, the shock dominates over the jet.  The angle at which the shock begins to compare to and/or dominate over the jet depends on the fluid velocity, and ranges from $5^{\circ}$ to $15^{\circ}$. }
    \label{fig:shock_energy}
\end{figure}

After the jet bores its way through the BNS ejecta and propagates through the outer material prescribed via Equation \ref{eq:density}, we find that a thin shock develops in front of the jet. This is apparent in panels  \textbf{(e)}, \textbf{(h)}, and \textbf{(k)} of Figure \ref{fig:hydro}, which show a thin shell with $\Gamma_{\infty} \sim 300$ in this region that becomes more pronounced at angles $\gtrsim 5^{\circ}$.  Zooming in on this region in Figure \ref{fig:hydro_shock} we see in panel  \textbf{(a)}, as indicated by the arrow, that the density far away from the jet axis increases very rapidly by multiple orders of magnitude just in front of the jet.  In panel \textbf{(b)} we see that this corresponds to a region of high temperature, which can lead to the up-scattering of photons to very high energies at these wide viewing angles.

We suspect this shock could be an early emergence of the external shock, caused by setting an un-physically high external density ($\rho_{\rm{ISM}}$ in Eq.~\ref{eq:density}) to ensure code stability. While this density was set in a way to ensure an external shock would not form in front of the core of the jet, the decrease of energy per solid angle away from the jet axis could justify an early onset of the shock. To test for that effect,
Figure \ref{fig:shock_energy} compares the energy content of different components of the jet,
\begin{equation}
    E_{jet} (\theta)  = \int_{0}^{\infty} dr \, r^{2} (T^{00} - \Gamma \rho c^{2} )
\end{equation}
to the energy required to produce a shock at radius $R$, which we calculate as
\begin{equation}
    E_{min} (\theta) = \frac{4 \, \pi}{3} \, R^3 \, \Gamma^2 \, c^2 \, \rho_{ISM},
    \label{eq:shock_energy}
\end{equation}
where $R$ is the radial coordinate of head of the jet, $\Gamma$ is the corresponding bulk Lorentz factor of the flow, and $\rho_{ISM}$ is the density floor used in Equation \ref{eq:density} to maintain code stability. We see that close to the jet axis the energy in the jet is an order of magnitude higher than this minimum energy and therefore on-axis observations are unaffected by an external shock.  Moving farther from the jet axis, $E_{jet}$ begins to decrease faster than $E_{min}$ until they intersect.  The angle at which $E_{jet}$ crosses $E_{min}$ is different for material that moves with different Lorentz factor.   If  we restrict to highly relativistic material, with $\Gamma \geq 10$, $E_{jet}$ decreases rapidly for $\theta > 5^{\circ}$ and crosses $E_{min}$ at $\theta=7.5^{\circ}$. If instead we include mildly relativistic material, such that $\Gamma \geq 2$, $E_{jet}$ decreases slowly, only crossing $E_{min}$ around $\theta = 15^{\circ}$. We conclude that light curves calculated at viewing angles larger than 5 degrees may be affected by the presence of the external shock.

\subsection{Emission}
In Figure \ref{fig:lightcurves} we plot the light curves obtained from mock observations conducted at viewing angles ranging from $\theta_{obs} = 1^{\circ}$ to $\theta_{obs} = 10^{\circ}$. Each panel has a solid red line representing observations where all photons within the above acceptance angle ranges are considered, and a blue dash-dotted line with photons from the external shock removed. For shallow viewing angles out to $\sim 4^{\circ}$, removing these photons amounts to removing part of the leading edge of the lightcurve, with the bulk of prominent peak being unaffected. For $\theta_{obs} \geq 6^{\circ}$, however, this leading edge constitutes the brightest part of the lightcurve. When keeping all photons, the peak luminosity initially decreases with viewing angle, but levels off around $6^{\circ}$.  However, with the shock removed, the peak luminosity steadily decreases with viewing angle. This agrees with our previous finding that an early development of the external shock is affecting the off-axis lightcurves.

It is interesting to note differences between lightcurves shown here and those of past studies. In contrast to the sharply peaked lightcurves on display here, those found in \cite{ito_global_2021} are much broader. Additionally, peak isotropic luminosities in \cite{ito_global_2021} are much higher than those obtained in this work. This is to be expected given the longer injection time of 4 s in \cite{ito_global_2021}, compared to 2 s here. 

Figure \ref{fig:spectra} shows time-integrated spectra corresponding to the lightcurves in Figure \ref{fig:lightcurves}, over the same range of viewing angles. The overall normalization varies with viewing angle analogously to the peak luminosity in Figure \ref{fig:lightcurves}, as expected.  Additionally, there is a steep drop-off in peak energy between $\theta_{obs} = 1^{\circ}$ and $\theta_{obs} = 2^{\circ}$, mirroring the angular structure of the jet shown in Figure \ref{fig:hydro}. Further steep changes in peak energy are also seen at higher viewing angles. It is worth noting that the peak frequencies seen here agree well with those found in \cite{ito_global_2021}. In both cases we see peaks of $\sim$ 1 MeV for on axis observations, which decrease to 10-20 keV when $\mathbf{\theta_{obs} \sim 3^{\circ}}$. The spectra in \cite{ito_global_2021} are, however, much more thermal
 than we find in this work.

\begin{figure*}
    \includegraphics[width=\textwidth]{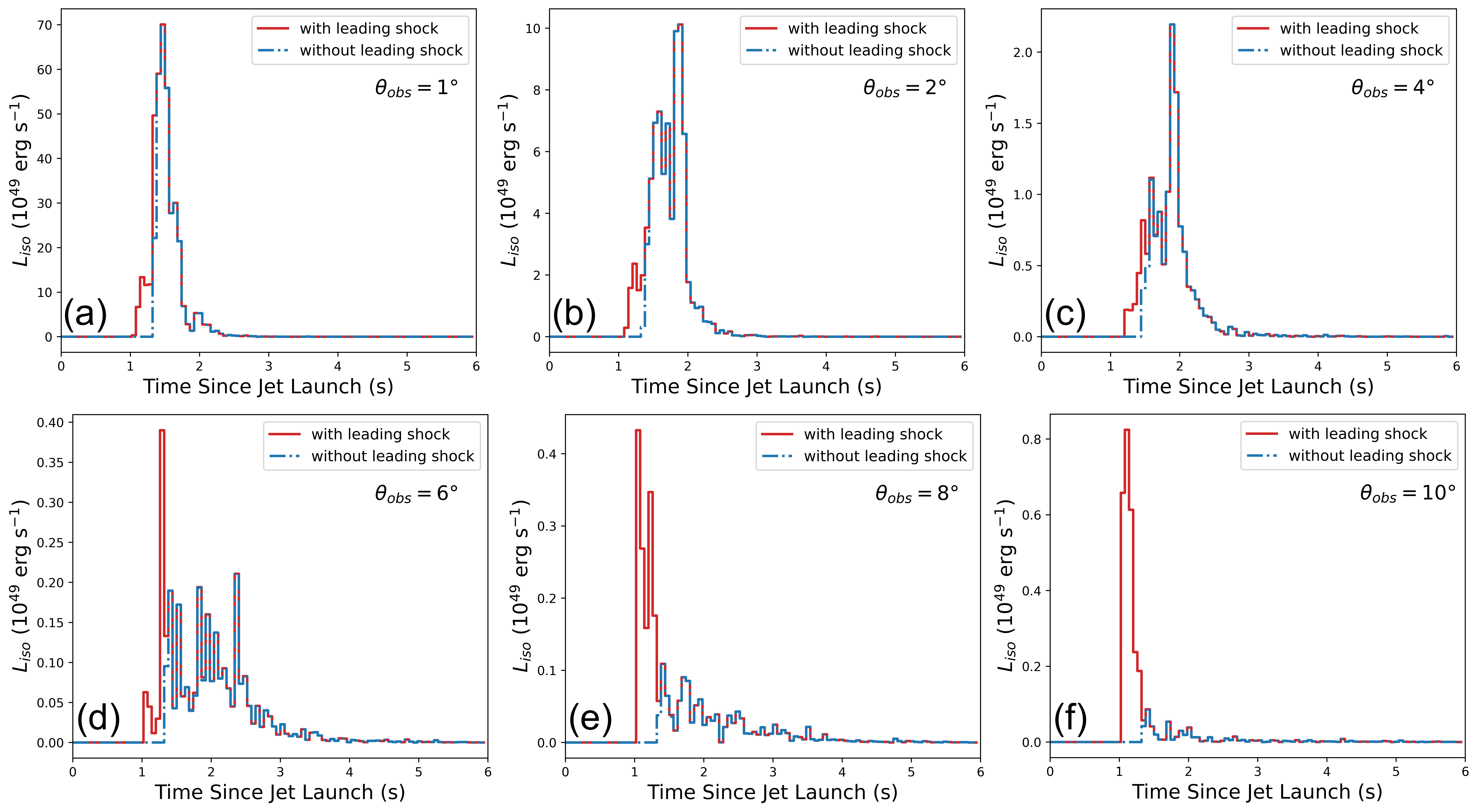}
    \caption{Bolometric lightcurves obtained from mock observations conducted at viewing angles ranging from $\theta_v = 1^{\circ}$ to $\theta_v = 10^{\circ}$. Note that jet injection lasts for 2 s, the first 1 s carried out in L21 and the final 1 s carried out in this work. Each panel shows a lightcurve comprising all photons selected during the mock observation (red line) alongside another lightcurve with photons from the thin shock (e.g. Figure \ref{fig:hydro} panel \textbf{(k.)}) removed (blue dashed line). The lightcurves shown in red maintain high peak luminosities at steep observing angles. The blue lightcurves, which remove the effects of an unphysical shock, exhibit the expected behavior of the peak luminosity steadily decreasing with observing angle.  All lightcurves were obtained by binning photon arrival times into 600 ms bins, to be consistent with \cite{poolakkil_fermi-gbm_2021}.}
    \label{fig:lightcurves}
\end{figure*}

\begin{figure*}
    \includegraphics[width=\textwidth]{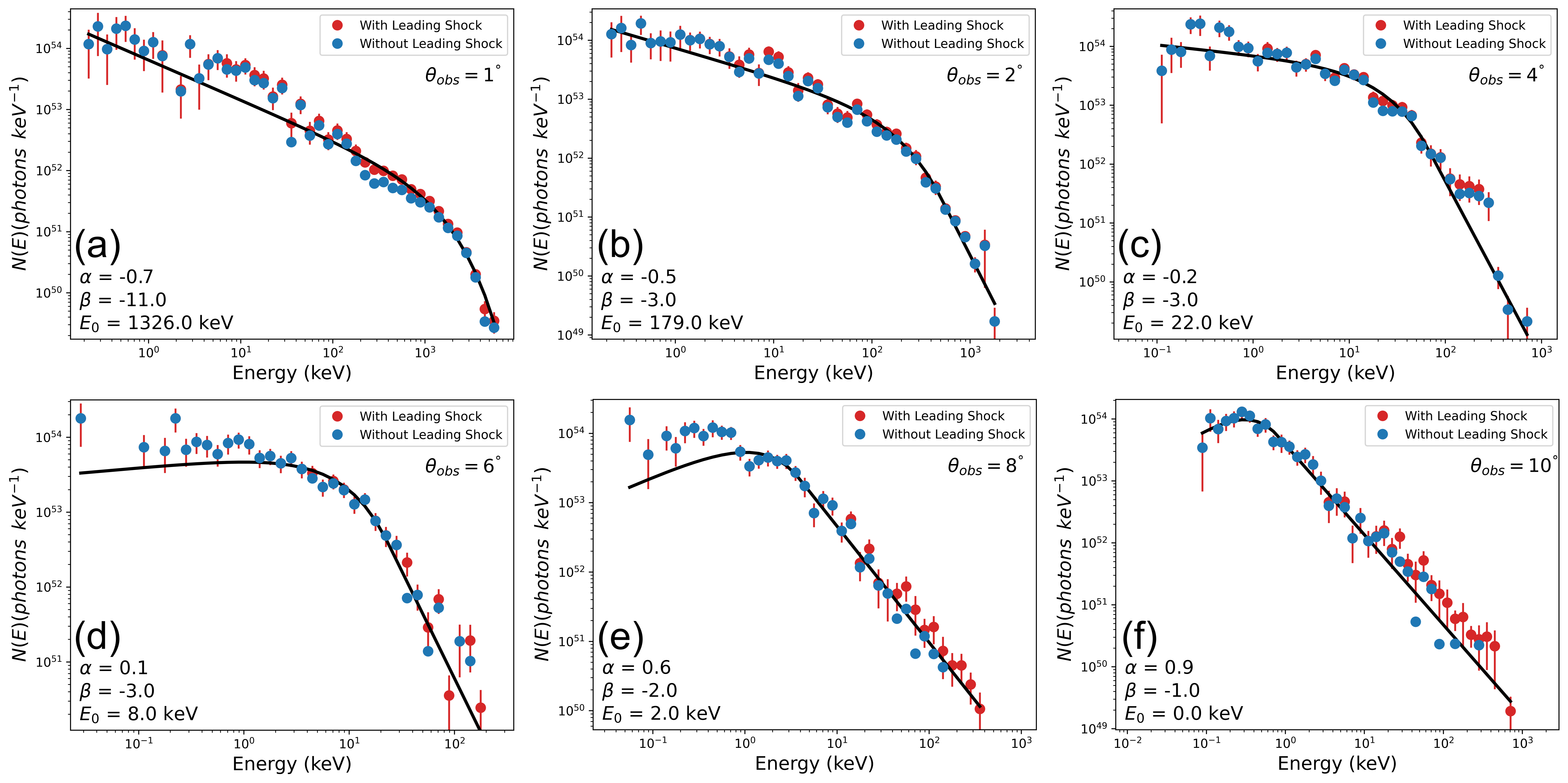}
    \caption{Time-integrated spectra corresponding to the lightcurves in Figure \ref{fig:lightcurves}. As with Figure \ref{fig:lightcurves}, the blue markers show spectra that include all photons, including those in the leading shock, while the blue markers show spectra with photons in the shock removed. The Band function is fit to the red points for each spectrum, obtained using the Markov-Chain Monte-Carlo algorithm implemented in \cite{foreman-mackey_emcee_2013}.  Peak energy is on the order of $1$ MeV for observations near the jet axis and decreases in tandem with the jet structure seen in Figure \ref{fig:hydro}. The low-energy slope parameter $\alpha$ is distinctly non-thermal in most spectra.}
    \label{fig:spectra}
\end{figure*}

Figures \ref{fig:lightcurves} and \ref{fig:spectra} show observations out to a viewing angle of $\theta_{obs} = 10^{\circ}$. To achieve this, we injected photons out to $20^{\circ}$ from the jet axis.  As the bulk Lorentz factor of the jet is small far from the jet axis, relativistic beaming effects are much less prominent than they are near the jet axis.  Thus, photons injected in the cocoon can be scattered at fairly large angles, as opposed to photons in the core of the jet which are tightly beamed with each successive scattering.  As such, in order to conduct a mock observation at high viewing angles, it is important to inject photons at angles significantly larger than the target $\theta_{obs}$. Any future study intending to construct models for highly off-axis emission would likewise have to inject photons even further than $20^{\circ}$.  In Figure \ref{fig:photon_positions} we show the jet density at $t = 12$ s, overlaid with the photon positions at that time. 

In Figure \ref{fig:correlations} we compare the observational Amati  correlation \citep{amati_intrinsic_2002} to our MCRaT results, using observational data from \cite{poolakkil_fermi-gbm_2021}. The red and blue points show values obtained from our MCRaT simulation, while the green and gray markers show real-world observations of SGRBs and LGRBs, respectively. Each data point from MCRaT is the result of a mock observation at a given observing angle.  Broadly speaking, the MCRaT simulations in this work lie closer to real-world observations of SGRBs than they do to LGRBs, particularly at shallow observing angles. There is, however, some strain at wider observing angles due to the sharp decreases in $E_{pk}$ at those angles. This pushes the corresponding mock observations into a region with no real-world observations. However, this region on the Amati plot coincides with a dark region for detectors where GRB detections are very unlikely. 

Figure \ref{fig:correlations} also distinguishes between observations in which photons in the external shock are included, to those observation in which those photons are removed. The former are shown in red while the latter are shown in blue. At shallow observing angles we find that our mock observations are virtually identical for both sets of photons, while at steep observing angles the two sets of points diverge.  $E_{pk}$ drops off rapidly at high viewing angles, resulting in a nearly vertical line at $E_{iso} \sim 10^{48}$ erg. This is most noticeable with the red points, and is mitigated with the blue points. Some observations result in spectra in which $\beta \geq -2$, which are shown in Figure \ref{fig:correlations} as hollow blue markers. For these observations, $E_{pk}$ of $\nu F_{\nu}$ isn't given by $(2 + \alpha) \, E_{0}$, as it is when $\beta \leq -2$. The actual peak energy for these points is likely higher than what is shown here.

Finally, in Figure \ref{fig:rad_eff} \textbf{(a)}, we plot the radiative efficiency, $ \eta ( \theta_{obs}) = E_{iso}/E^{k}_{iso}$ for each mock observation. $E_{iso}$ is the isotropic equivalent luminosity of an MCRaT observation conducted at observing angle $\theta_{obs}$ and $E^{k}_{iso}$ is the isotropic equivalent kinetic energy of the jet along the line-of-sight at $\theta_{obs}$. $E_{iso}$ is obtained as it was in Figure \ref{fig:correlations}, by integrating over the lightcurve of an observation for a particular $\theta_{obs}$, while $E^{k}_{iso}$ is obtained by integrating the energy density of the jet ($T^{00}$ from equations \ref{eq:cons} and \ref{eq:stress}), subtracted by the rest energy density: 
\begin{equation}
\begin{split}
    E^{k}_{iso} &= \int d^3x \, (T^{00} -\Gamma \rho c^2 )\\ &= 4\pi \int dr \, r^2 \, \left\{ \left( \rho c^2 + 4p \right)  \, \Gamma^2- p - \Gamma \rho c^2\right\},
\end{split}
\label{eq:kinetic_energy}
\end{equation} 
along the line-of-sight at a particular angle $\theta$, measured from the jet axis. Each mock observation conducted at observing angle $\theta_{obs}$ collects photons whose position in the jet, \textit{before} ProcessMCRaT simulates a detection, can vary over some range $\theta_{p}$. We compute Equation \ref{eq:kinetic_energy} for each value of $\theta_p$ found in an observation and compute a weighted average with respect to the number of photons found at each $\theta_p$. Figure \ref{fig:rad_eff} shows $\eta(\theta_{obs})$, with blue diamonds showing $\eta$ where $E_{k}$ is integrated over the entire jet, orange squares showing only relativistic material with $\Gamma \geq 2$, and green circles likewise selecting for $\Gamma \geq 5$. For $\Gamma \geq 1$ and $\Gamma \geq 2$,  $\eta$ is no larger than a few percent. When including only highly relativistic material, however, $\eta$ can approach 20\% for off-axis observations. We find that the efficiency of relativistic material is generally higher than when we consider the entire jet. Regardless of which components of the jet we integrate over, $\eta$ is considerably smaller than what has been seen in previous photospheric models of GRB prompt emission \citep{lazzati_high-efficiency_2011}, and can vary by an order of magnitude over the $10^{\circ}$ observing angle range considered in this work.

In Figure \ref{fig:rad_eff} \textbf{(b)} we plot $T_{90} = t_{95}-t_{5}$ obtained from each mock observation, where $t_{x}$ is defined by
\begin{equation}
    \frac{E(t_{x})}{E_{tot}} = \dfrac{\int_{0}^{t_{x}} dt \, L_{iso} (t)}{\int_{0}^{\infty} dt \, L_{iso} (t)}  = \frac{x}{100},
\end{equation}
with $x/100$ denoting the fraction of energy contained in the lightcurve between $t=0$ and $t=t_x$. The red circles show $T_{90}$ for observations that include photons in the external shock, while the triangles show $T_{90}$ with photons in the external shock removed.  Regardless of which set of photons we consider most observations satisfy the condition $T_{90} \leq 2$ s for SGRBs.  However, the few with $T_{90} \geq 2$ s indicate the possibility that some BNS mergers could produce low-luminosity bursts that don't classify as SGRBs according to the standard $T_{90} \leq 2$ s cutoff.

\begin{figure}
    \centering
    \includegraphics[width=\columnwidth]{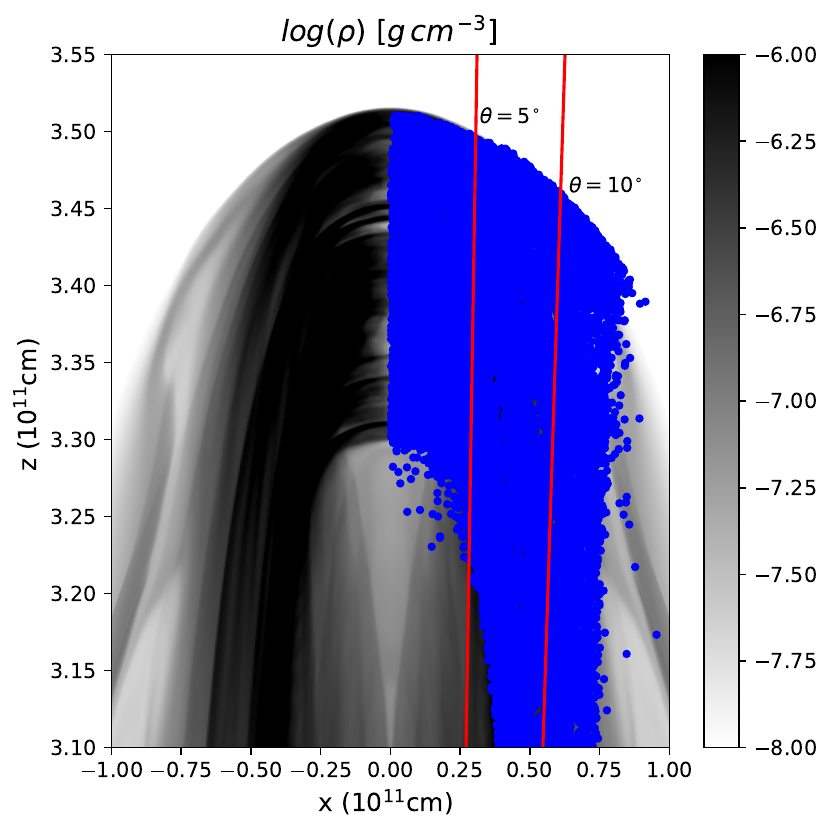}
    \caption{Jet density at simulation time $t = 12$ s, overlaid with photon positions (blue dots). Photons propagate with the jet, within a region of $\lesssim 3 \times 10^{10}$ cm in length, and within the first $5^{\circ}$ of the jet axis.  Between 5 and 10$^{\circ}$, photons begin to diffuse radially across a much lager area. For  $\theta > 10^{\circ}$, photons diffuse even more, which affects the statistics of producing lightcurves and spectra.}
    \label{fig:photon_positions}
\end{figure}

\begin{figure}
    \includegraphics[width=\columnwidth]{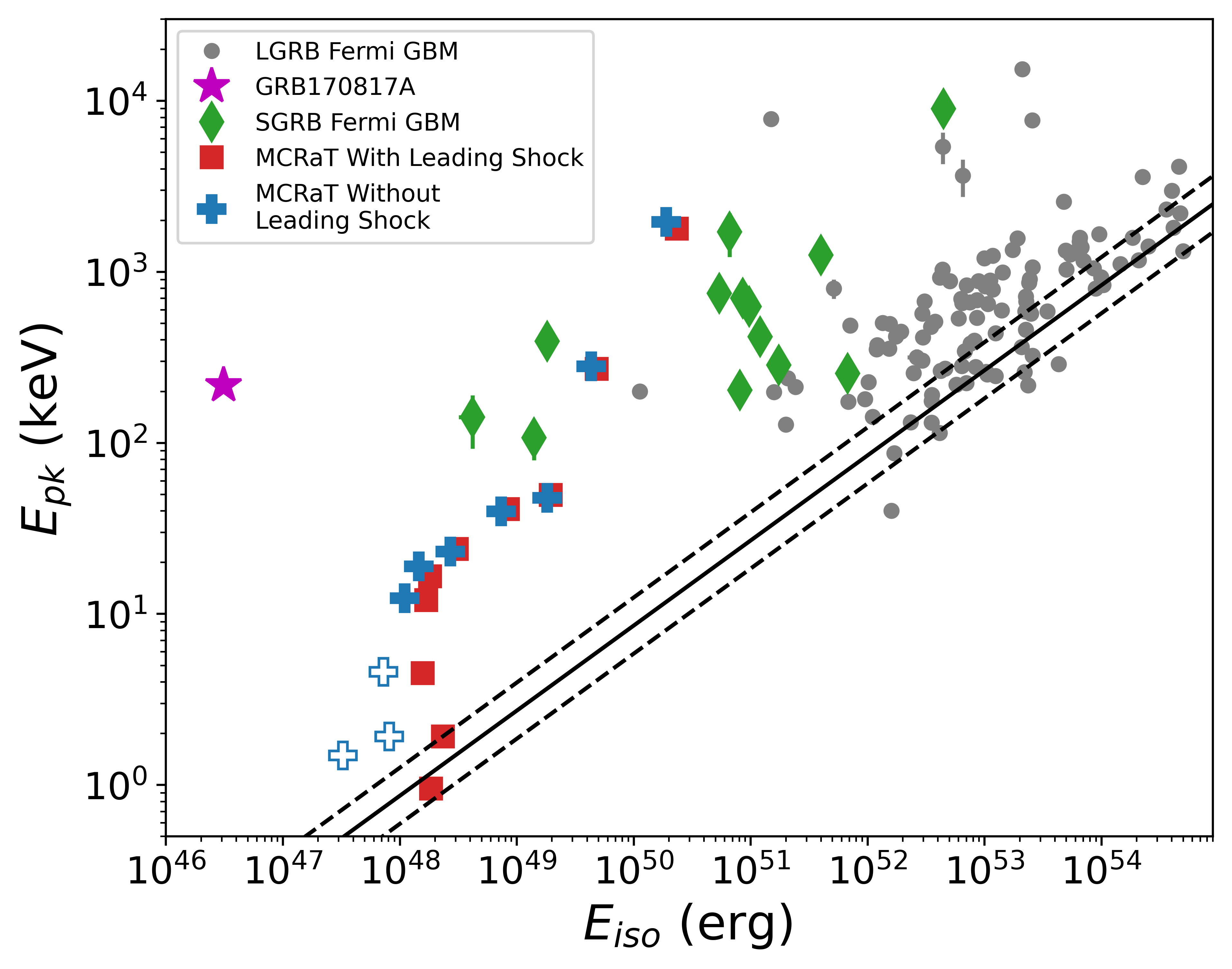}
    \caption{The Amati correlation.  Red squares denote data points obtained from MCRaT simulations, while gray circles and green diamonds denote observational data obtained from \cite{poolakkil_fermi-gbm_2021}.  $E_{pk}$ and $E_{iso}$ were obtained by fitting the Band function (equation \ref{eq:band}) to each spectrum.  Green diamonds denote SGRB observations with $T_{90} <2$ s cutoff, while gray circles denote LGRBs with a $T_{90} \geq 2$ cutoff. Blue crosses show MCRaT observations with photons from the leading shock removed. Hollow blue crosses denote observations with $\beta \geq -2$.}
    \label{fig:correlations}
\end{figure}

\begin{figure}
    \includegraphics[width=\columnwidth]{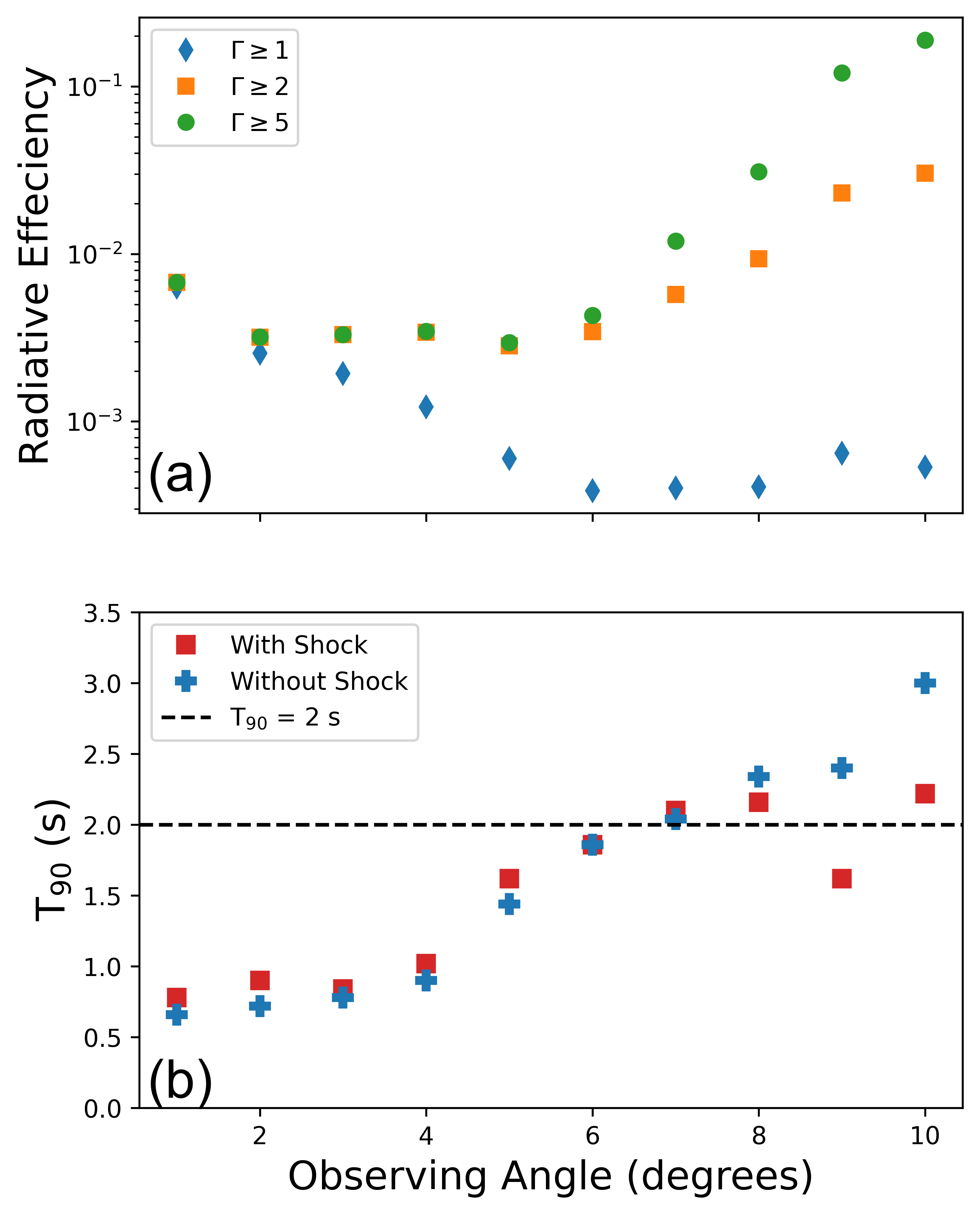}
    \caption{\textbf{(a)} The radiative efficiency $\eta = E_{iso}/E_{kin}$ of MCRaT photon spectra, where $E_{iso}$ is obtained from mock observations and $E_{k}$ is the kinetic energy of the jet, integrated over the line of sight.  Each point is obtained from a mock observation at the given observing angle.  When integrating $E_{k}$ over the entire line of sight (blue diamonds), the radiative efficiency decreases by an order of magnitude from  $ \theta_{obs} = 1^{\circ}$ to $10^{\circ}$. If we only integrate over relativistic material (green circles) the radiative efficiency can still vary dramatically, but increases at high viewing angles. \textbf{(b)} $T_{90}$ computed by integrating over  lightcurves in Figure \ref{fig:lightcurves}, corresponding to the same observing angles in panel \textbf{(a)} of this figure. Note that the jet injection lasts for 2 s. While most mock observations satisfy the condition $T_{90} \leq 2$ s for short-duration bursts, there are some instances where $T_{90} > 2$ s, indicating a possibility for BNS mergers to produce low-luminosity bursts that won't be strictly categorized as an SGRB.  }
    \label{fig:rad_eff}
\end{figure}

\section{Summary and Discussion} 

In this paper we model the photospheric prompt emission of an SGRB. While there has been previous work approaching the same fundamental problem \citep{ito_global_2021}, the work presented here is the final step of the first end-to-end simulation that began with a 3D GRMHD simulation of a BNS merger carried out by \citealt{ciolfi_first_2019}, and ends with the prompt emission of an SGRB. The BNS merger carried out in \citealt{ciolfi_first_2019} was followed by a 3D RHD simulation of a relativistic jet injected into the BNS merger ejecta, and lasted until jet breakout (L21). Here, we continue that simulation in 2D until the jet approaches the photosphere at $\sim 10^{13}$ cm.  Then we use the MCRaT radiation transfer code to model the resulting photospheric emission. 

We find that the jet, which develops a complicated radial structure and cocoon, maintains the initial collimation inherited from L21.  We also observe that a thin shock develops in front of the jet over the course of the RHD simulation.  This shock, which is essentially a premature afterglow, develops due to the value for $\rho_{ISM}$ adopted in this work being orders of magnitude higher than typical for the inter-stellar medium, and is more prominent far from the jet axis. Off-axis observations that include photons that interact with this part of the jet can thus carry fingerprints of the shock when its energy becomes comparable to that of the relativistic jet. However, we are generally able to separate the effects of the thin shock from those of the relativistic jet proper. While similar effects could arise from processes such as a cocoon break-out \citep{lazzati_off-axis_2017}, our static treatment of BNS merger ejecta would cause this to happen too early to be included in our radiative transfer.

Lightcurves obtained from MCRaT calculations show bursts that last for $\sim 1$ s, characteristic of SGRBs. However, the light curve duration increases off-axis with some observations resulting in $T_{90} \sim 3$ s, thus opening the door for possible low-luminosity bursts that wouldn't normally be associated with BNS mergers.  We note that the possibility of producing bursts of this character, which would traditionally be considered LGRBS, from binary
neutron stars mergers has received considerable attention recently,
due to the discovery of optical and infrared kilonova signatures associated with two GRBs lasting approximately 10 seconds \citep{Rastinejad2022, Troja2022, Yang2022, Levan2023}. While these bursts are both too long and too bright to be explained by the jet parameters we adopted, it is of high value to understand more generally the conditions under which traditionally long GRBs can be produced by a BNS merger. Here, we provide an interpretation of the role that GRBs with $T_{90} \sim 3$ s in this work play within this larger context.

The diversity of prompt emission durations from compact binary mergers GRBs, extending into what has been traditionally considered the realm of long GRBs, has been interpreted via a unification model by \citet{Gottlieb2023,
Gottlieb2025}. This model suggests that the distinction between short and long  GRBs from binary compact object mergers depends 
 on both the nature of the compact remnant and the post-merger disk mass
left behind to accrete. Within this framework, the newly discovered class of
long GRBs emerges from black hole remnants surrounded by a massive 
($M_d\gtrsim 0.1 M_\odot$) post-merger accretion disk.

Because this unification scenario is highly sensitive to both the remnant type and the post-merger disk mass, it offers a novel avenue for probing the nuclear physics of neutron star matter. Specifically, \citet{Perna2025} highlighted that the ratio of long to short GRBs resulting from binary neutron star mergers is strongly dependent on a critical parameter: the threshold mass that determines whether a neutron star remnant is short- or long-lived. To derive meaningful constraints from this framework, accurate classification of observed GRBs originating from compact object mergers is essential.

The long-duration GRBs identified in this work are characterized by low isotropic energies ($E_{\rm iso}\sim 10^{48}$~erg), typically below those of standard short GRBs detected by Fermi-GBM. However, since GRBs with kilonova counterparts tend to occur at closer distances than typical short GRBs, such low-luminosity events are more likely to be included in observational samples. For these GRBs, a key question is whether their longer durations are genuinely linked to the nature of the merger remnant, or are instead a consequence of larger viewing angles. A crucial diagnostic in this context is the combination of low isotropic energies and high observing angles.

We note also that lightcurves are affected by shock forming in front of the jet. While the peak luminosity would be expected to decrease with viewing angle, we find that it approaches a roughly constant, minimum value of $\sim 5 \times 10^{48}$~erg s$^{-1}$ when $\theta_{obs}$ approaches $10^{\circ}$. Removing photons from this region results in lightcurves that are more consistent with behavior expected from observational correlations.

The accompanying MCRaT spectra peak around $\sim 1$ MeV for $\theta_{obs} \sim 1^{\circ}$ and drops off when $\theta_{obs} \gtrsim 5^{\circ}$. We also find that spectra obtained from on-axis observations have distinctly non-thermal low-energy tails, with the low energy Band parameter ranging from $\alpha \sim -0.7$ to $-0.2$.  Photospheric models of GRBs are typically characterized by hard spectra, particularly at low energies, which is in stark contrast to what we find in this work.  This is likely due to the hydrodynamical variables which vary on length scales smaller than the jet (e.g. Figure \ref{fig:hydro}).  In previous studies, typically on LGRBs, jets vary much more smoothly on length scales  comparable to those that photons travel while scattering with the jet.  Here, the jet energy varies strongly, even deep in the core of the jet.  Photons thus have to scatter through not just the fast, hot material, but also through much slower and cold material.  The result would be a jet that is less efficient at up-scattering photons, which would translate to more low energy photons, and thus  softer spectra. Similar results were found in \cite{parsotan_photospheric_2018} with LGRBs.

We compare our simulations to observational correlations of real GRBs and find that our work is broadly consistent with the Amati correlation. For slightly off-axis observations, and to a lesser extent in on-axis observations, variation in viewing angle is able to reproduce the slope in the Amati correlation, although this is strained in off-axis observations. However, these points lie in a region where real-world observations are unlikely with current observational capabilities.  Additionally, what strain there is at high viewing angles is largely due to the significant drops in peak energy in those observations. Given the steep high energy tail of these observations ($\beta \geq -2$), it is possible that the true $E_{pk}$ from these observations is in reality much higher than what we find in this work. Fitting our spectra to the $8 \,\text{keV} - 40 \,\text{MeV}$ energy range corresponding to the Fermi detector would likely help reduce $\beta$ and result in more reliable spectral peaks. Additionally, a previous study \citep{walker_role_2024} modeled the role a neutron component would have on the prompt emission of GRBs, showing that the overall effect would be to increase both the peak energy and luminosity of GRBs. While the present work doesn't take a neutron component into account, BNS mergers would naturally contribute a sizable neutron component in a baryon loaded jet, which could result in better agreement with correlations.

Given the wide range of peak and isotropic energies exhibited in Figure \ref{fig:correlations}, it is interesting to directly compare the prompt emission of GRB170817A (purple star) to our results. The lightcurve and spectrum of GRB170817A have been challenging to characterize, largely due to a borderline duration for a GRB ($T_{90}\sim2$~s, \citealt{goldstein_ordinary_2017}) and a
low energy budget ($\sim10^{47}$~erg), as described above. Together with the fact that GRB170817A does not
fit in the Amati or Yonetoku  relations \citep{Pozanenko18}, the origin of the prompt spectrum of GRB170817A/GW170817 has been highly debated. Explanations for a departure from those relations typically require either a different emission mechanism, such as a relativistic shock breakout \citep{kasliwal_illuminating_2017}, or the effect of an off-axis geometry \citep{Mooley2018}, resulting in a much higher peak frequency than expected for such a low-luminosity event. We note that our synthetic off-axis observations (blue crosses in Figure \ref{fig:correlations}) have energy comparable to GRB170817A, along with large high-frequency spectral indices, hinting at peak frequencies of $\sim 100$ keV or more. Pinning down more precise values for the peak energies of these mock observations will likely require fitting them in the range corresponding to the Fermi detector.

Finally, we compute the radiative efficiency of our models at different observation angles, and find that the observed radiative efficiency can vary with $\theta_{obs}$ by more than an order of magnitude. This further underscores the importance of the dependence GRB emission has on observing angle. We also find that the radiative efficiency is much lower than previous studies of photospheric emission of long-duration GRBs (e.g. \citealt{lazzati_high-efficiency_2011}), which is in line with previous estimates of the observed radiative efficiency of GRB 170817A \citep{salafia_accretion--jet_2021}. In particular, we find efficiencies no greater than a few percent, and in some cases as low as $\sim$ 10$^{-3}$ , much lower than the efficiencies of $\sim 90 \%$ seen in previous works. These low efficiencies, which result from the hydrodynamical behavior of the jet, suggest that the non-thermal spectra can result from inefficient radiative processes. While more work needs to be done to investigate the processes that shape the maximum radiative efficiency of systems that produce GRBs, these results hint at the ways in which the spectral parameters of photospheric models can be associated with the physical constraints of the processes that can produce relativistic jets. 

It's worth noting that, while we simulate the jet out to $10^{13}$ cm, the radiation is still somewhat coupled to the outflow, especially at large observation angles.  This is a necessary trade off for maintaining high resolution throughout the simulation, which is in contrast to \cite{ito_global_2021} where the authors sacrifice resolution far away from the central engine in order to simulate the jet out to $10^{14}$ cm. Allowing the jet to travel far enough for the radiation to more fully decouple from the jet would likely result in colder spectra due to the cooling of the jet that would occur from $10^{13}$ cm to $10^{14}$ cm.  \cite{walker_role_2024} showed that GRB spectra produced with MCRaT evolve from thermal to non-thermal and, if that trend holds, here the non-thermal low energy spectra would be maintained throughout longer simulations.  However, simulating jets until their radiation can fully decouple, while maintaining high resolution is an important next step that could help elucidate some of the open questions that still remain. 

% The results presented here are very promising, as they expand the parameter space of photospheric models of GRBs. Previously, thermal low energy spectra that emerge from photospheric models have been a hurdle to agreeing with observations, and here we show that significantly non-thermal spectra are possible. While we don't reproduce GRB170817A, we show that photospheric models are consistent with observational correlations of SGRBs.  While there is room to make our models more accurate, it is evident that photospheric models of GRB prompt emission are an excellent candidate for explaining GRB emission.  

%% IMPORTANT! The old "\acknowledgment" command has be depreciated. It was
%% not robust enough to handle our new dual anonymous review requirements and
%% thus been replaced with the acknowledgment environment. If you try to 
%% compile with \acknowledgment you will get an error print to the screen
%% and in the compiled pdf.
%% 
%% Also note that the akcnowlodgment environment does not support long amounts of text. If you have a lot of people and institutions to acknowledge, do not use this command. Instead, create a new \section{Acknowledgments}.

\section{Acknowledgments}
N.W. and D.L. acknowledge support from NSF grant AST-1907955.  
R.P. acknowledges support from NSF award AST-2006839 and NASA award 80NSSC25K7554.
Resources supporting this work were provided by the NASA High-End Computing (HEC) Program, through the NASA Advanced Supercomputing Division at Ames Research Center.

The authors acknowledge the Texas Advanced Computing Center (TACC) at The University of Texas at Austin for providing computational resources that have contributed to the research results reported within this paper.

The authors acknowledge the Advanced Research Computing Services (ARCS @OSU) supercomputing resources (https://arcs.oregonstate.edu) made available for conducting the research reported in this paper.

%% To help institutions obtain information on the effectiveness of their 
%% telescopes the AAS Journals has created a group of keywords for telescope 
%% facilities.
%
%% Following the acknowledgments section, use the following syntax and the
%% \facility{} or \facilities{} macros to list the keywords of facilities used 
%% in the research for the paper.  Each keyword is check against the master 
%% list during copy editing.  Individual instruments can be provided in 
%% parentheses, after the keyword, but they are not verified.

\vspace{5mm}
%% Similar to \facility{}, there is the optional \software command to allow 
%% authors a place to specify which programs were used during the creation of 
%% the manuscript. Authors should list each code and include either a
%% citation or url to the code inside ()s when available.

\software{MCRaT \citep{lazzati_monte_2016},  
          ProcessMCRaT \citep{parsotan_parsotatprocessmcrat_2021}, 
          matplotlib \citep{hunter_matplotlib_2007},
          emcee \citep{foreman-mackey_emcee_2013}
          }

%% Appendix material should be preceded with a single \appendix command.
%% There should be a \section command for each appendix. Mark appendix
%% subsections with the same markup you use in the main body of the paper.

%% Each Appendix (indicated with \section) will be lettered A, B, C, etc.
%% The equation counter will reset when it encounters the \appendix
%% command and will number appendix equations (A1), (A2), etc. The
%% Figure and Table counter will not reset.

%% For this sample we use BibTeX plus aasjournals.bst to generate the
%% the bibliography. The sample631.bib file was populated from ADS. To
%% get the citations to show in the compiled file do the following:
%%
%% pdflatex sample631.tex
%% bibtext sample631
%% pdflatex sample631.tex
%% pdflatex sample631.tex

\bibliography{MyLibrary}{}
\bibliographystyle{aasjournal}

%% This command is needed to show the entire author+affiliation list when
%% the collaboration and author truncation commands are used.  It has to
%% go at the end of the manuscript.
%\allauthors

%% Include this line if you are using the \added, \replaced, \deleted
%% commands to see a summary list of all changes at the end of the article.
%\listofchanges

\end{document}